\documentclass[manuscript]{aastex}

\shorttitle{The Tamara family}
\shortauthors{Novakovi\' c et al.}

\begin{document}


\title{A dark asteroid family in the Phocaea region}


\author{Bojan~Novakovi\'c}
\affil{Department of Astronomy, Faculty of Mathematics, University of Belgrade, \\
Studenski trg 16, 11000 Belgrade, Serbia}
\email{bojan@matf.bg.ac.rs}

\author{Georgios Tsirvoulis}
\affil{Astronomical Observatory, Volgina 7, 11060 Belgrade 38, Serbia}

\author{Mikael Granvik}
\affil{Department of Physics, P.O. Box 64, 00014 University of Helsinki, Finland}

\and

\author{Ana Todovi\' c}
\affil{Department of Astronomy, Faculty of Mathematics, University of Belgrade, \\
Studenski trg 16, 11000 Belgrade, Serbia}

\begin{abstract}
We report the discovery of a new asteroid family among the dark
asteroids residing in the Phocaea region the Tamara family. We make use of available
physical data to separate asteroids in the region according to their
surface reflectance properties, and establish the membership of the family. We
determine the slope of the cumulative magnitude distribution of the
family, and find it to be significantly steeper than the corresponding
slope of all the asteroids in the Phocaea region.  This implies that sub-kilometer dark
Phocaeas are comparable in number to bright $S$-type objects, shedding
light on an entirely new aspect of the composition of small Phocaea
asteroids.  We then use the Yarkovsky V-shape based method and
estimate the age of the family to be $264\pm43$~Myr.  Finally, we
carry out numerical simulations of the dynamical evolution of the
Tamara family. The results suggest that up to 50 Tamara members with absolute
magnitude $H<19.4$ may currently be found in the near-Earth region.
Despite their relatively small number in the near-Earth space, the 
rate of Earth impacts by small, dark Phocaeas is non-negligible.
\end{abstract}

\keywords{celestial mechanics --- minor planets, asteroids: general}

\section{Introduction}

Understanding the structure and past evolution of the asteroid families in the inner
asteroid belt are important for constraining the history and
evolution of the belt, as well as the delivery of asteroids to the
near-Earth region \citep{bottke2015,mikael2017}.

The Phocaea region is a high-orbital-inclination part of the inner
asteroid belt which is separated from the low-inclination asteroids by
the $\nu_6$ secular resonance. The region has dynamical boundaries
from all sides but one, making it almost completely detached from the
rest of the asteroid belt \citep{knemil2003,michtchenko2010}.

Most of the Phocaeas\footnote{We use term \textit{Phocaeas} to refer to all asteroids
from the Phocaea region, while possible additional restrictions are always
explicitly mentioned.} are classified as $S$-type asteroids
\citep{carvano2001}, typical for objects in the inner asteroid
belt. Still, the relative mass contribution of each taxonomic class
changes with size in each part of the asteroid belt \citep{DeMeo2014}.
The most obvious correlation is an increase of $C$-type objects as
size decreases in the inner belt. Nevertheless, as far as the total
number of asteroids that belong to a specific spectral type is concerned, the Phocaea region
has been thought to be dominated by bright, $S$-type objects \citep{carvano,wise}.

A large fraction of asteroids from the Phocaea region belong to the
Phocaea collisional family \citep[e.g.][]{milani2014}, estimated to be
about 1.2~Gyr old \citep{milani2016}. The possible existence of other
families inside this region has been discussed by several authors
\citep{gilhutton2006,carruba2009,nov2011,masiero2013} who proposed
several candidate groups which might be collisional families. Still,
no family formed by a break-up of a dark carbonaceous parent body has
been proposed to exist in the Phocaea region. However, as we show
later in this paper, there are strong indications that such family
exists.  For this reason, we focused on the population of dark
($C$-type) asteroids located in the region, and searched for potential
tracers of a collisional family among these objects.

\section{Dark Phocaea asteroids: identification and search for a family}

There are currently 4,072 known numbered and multi-opposition asteroids in the Phocaea
region\footnote{Data obtained from the AstDyS service
  (hamilton.dm.unipi.it/astdys/). We define the Phocaea region using
  the following ranges in proper orbital elements: $2.1<a_p<2.5$~AU,
  $0.0<e_p<0.4$, and $0.3<sin(i_p)<0.5$.}. Using physical data
obtained by the Wide-field Infrared Survey Explorer
\citep[WISE;][]{wise}, we found that the region consists mostly of
bright asteroids ($\sim73\%$) with albedos higher than $0.1$.

\begin{figure}
\begin{center}
 \includegraphics[width=0.7\textwidth,angle=-90]{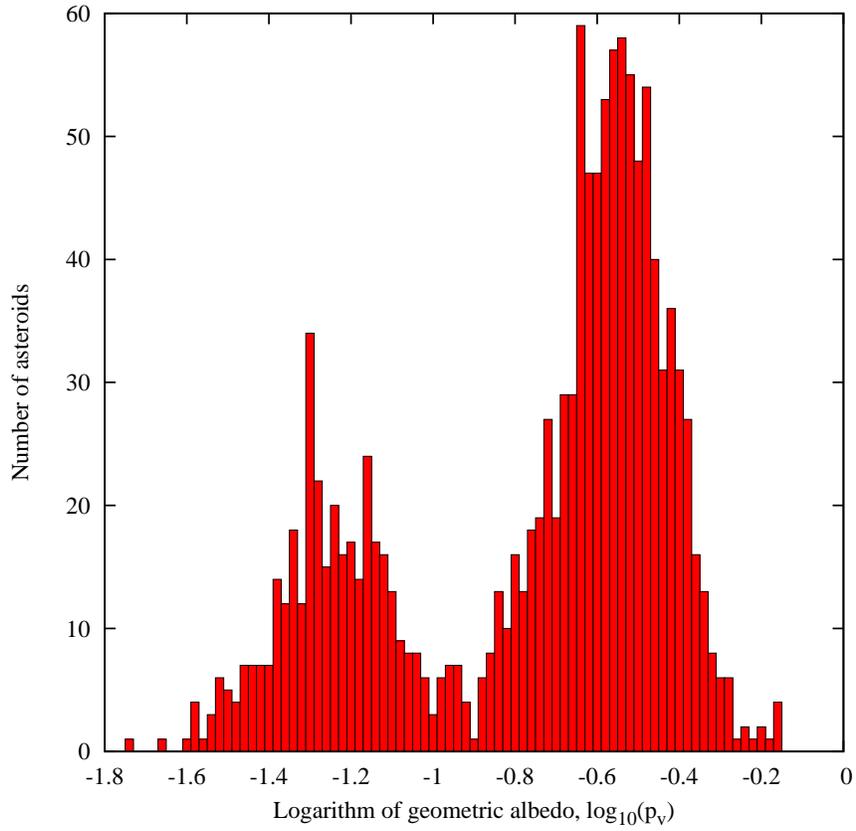}
 \caption{Geometric albedo distribution of asteroids in the Phocaea
   region. The presence of two distinct sub-populations is apparent.
   The left and right peak in the distribution represent dark and 
   bright objects respectively.}
 \label{Fig:hist}
 \end{center}
\end{figure}

However, the albedo distribution shows a clear separation between dark
and bright asteroids (Figure~\ref{Fig:hist}). It is roughly the sum of
two separated Gaussians, and the spreading of the low albedo part is
narrower than what we usually observe across the asteroid belt. This
suggests the possible existence of a dark asteroid family.

The number density of dark asteroids in the region is far lower than
the total number density, making the possible dark family totally
indistinguishable if one looks at the whole asteroid population. The
only way to study this family is to consider solely dark
asteroids.\footnote{A similar effort was done by \citet{masiero2013},
  who studied separately the low and high albedo asteroids across the
  main belt. However, they failed to identify this group due to the
  significant overlap between their two albedo populations.}

\subsection{The identification of dark asteroids}

The next step in our study was to obtain a catalog of dark asteroids
in the region. Following \citet{walsh2013} we adopted to work with
objects having geometric albedo $p_v$ below $0.1$. The WISE data
provides albedos for 1,280 out of the 4,072 asteroids in the region,
and of those 1,280 we found 348 dark ones. In an effort to expand this
catalog, we selected, in a similar manner, dark asteroids as
identified by the AKARI \citep{akari}, and the Infrared Astronomical
Satellite (IRAS) \citep{iras} surveys, where we found 41 and 12
low-albedo asteroids, respectively.

We also made use of the MOVIS catalog \citep{movis}, which uses VISTA
colors in order to distinguish between $C$- and $S$-complex
asteroids. According to \citet{movis} the (Y-J) vs (Y-Ks) color space
provides the largest separation between the two complexes, the
separatrix being the line
$(Y-J)=0.338^{\pm0.027}\cdot(Y-Ks)+0.075^{\pm0.02}$. Therefore we
considered as $C$-type those asteroids whose entire 1-$\sigma$ error
bar lies below this line. This way we obtained $8$ dark asteroids.

Finally we extracted dark asteroids as characterized by
\citet{carvano} using the Sloan Digital Sky Survey (SDSS) data
\citep{sdss}. There are 76 objects classified as either $C$- or
$D$-type, but for the purpose of this work we used only 16 asteroids
which have $>50\%$ probability to belong to the one of specified taxonomic
types.

In this way we identified 381 dark objects, and after removing 5
asteroids with contradictory albedos\footnote{In our sample asteroids
  (587), (2105), (4899), (8356), and (74749) have albedo determined
  from two different surveys but the results are inconsistent.}, we
obtained the catalog of dark Phocaeas containing 376 objects.

\begin{figure}
\begin{center}
 \includegraphics[width=0.72\textwidth,angle=-90]{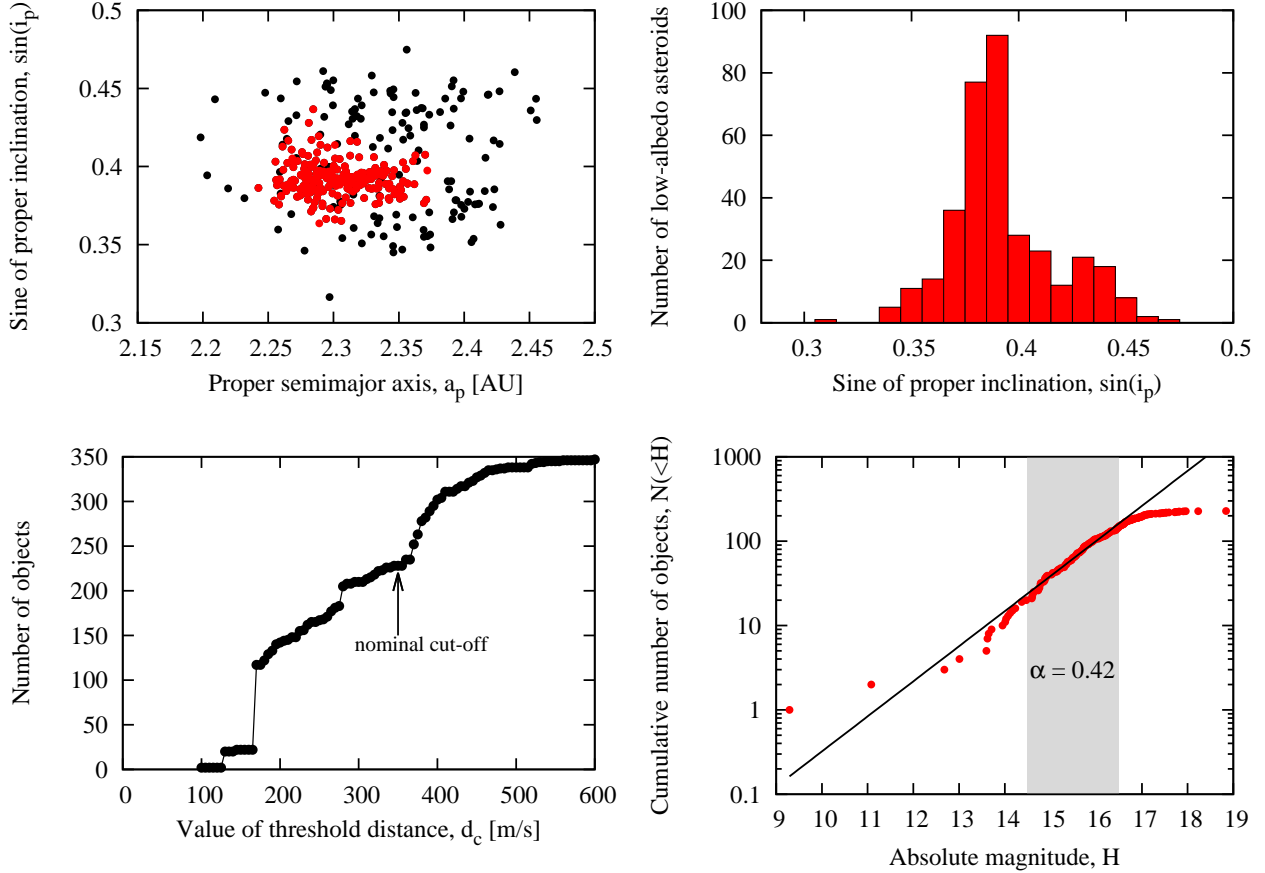}
 \caption{Properties of low-albedo asteroids in the Phocaea
   region. \textit{Top-left panel:} Proper semimajor axis versus sine
   of proper inclination projection of the dark asteroids in the
   Phocaea region (black dots), and Tamara family members (red dots);
   \textit{Top-right panel:} Histogram of orbital inclinations of dark
   Phocaeas; \textit{Bottom-left panel:} Number of asteroids
   associated with the family as a function of cut-off velocity;
   \textit{Bottom-right panel:} The cumulative magnitude distribution
   of the Tamara family (red dots), and its the best-fitting line
   (black solid line). The dashed area indicates the interval used for
   the fitting. }
 \label{Fig:4panels}
 \end{center}
\end{figure}

The dark component of this population has a non-uniform number density
in the proper elements space (Figure~\ref{Fig:4panels}), suggesting
the presence of an asteroid family.

\subsection{The search for a family}

We performed Hierarchical Clustering analysis \citep{hcm1990} on the
catalog of dark objects to obtain the membership of the new family. We
selected the asteroid (326)~Tamara as the starting body, because it is
potentially the largest member of the family, and increased the cutoff
velocity from $100$ to $600$~m/s in steps of $5$~m/s. The cutoff
velocities are overall higher than what we usually encounter in
similar studies, but the fact that we use only dark asteroids whose
number-density is low in this region justifies these
values.\footnote{\citet{nes2015} used $150$~m/s to determine the
  membership of the Phocaea family within the whole population of
  Phocaeas. As known dark asteroids account for about 9\% of all
  Phocaeas (376/4072), but occupy almost the same volume in the
  orbital space, a reasonable value to define the dark family should
  be about $\sqrt{11}$ times larger than the one used for the Phocaea
  family. This is why we select $350$~m/s as the nominal cut-off,
  rather than a value from the first plateau seen around $250$~m/s.}

The result is shown in the bottom-left panel of
Figure~\ref{Fig:4panels}. The family membership is defined at the
cutoff of $350$~m/s, because this value belongs to the well defined
\textit{plateau} visible in Figure~\ref{Fig:4panels} \citep[see
  e.g.][for details on this methodology]{nov2011}.  The nominal cutoff
corresponds to 226 members, which equals about 60\% of all dark
asteroids found in the region. This membership includes asteroid
(326)~Tamara\footnote{Despite being linked in some classifications to
  the Phocaea family, with an albedo of $0.040\pm0.002$ \citep{akari},
  the asteroid Tamara obviously does not belong to this group.}, as
the largest family member. Therefore, we named this group the Tamara
family.

\begin{deluxetable}{rrrrrrrr}
\tabletypesize{\scriptsize}
\tablecaption{List of the Tamara family members}
\tablewidth{0pt}
\tablehead{
\colhead{Name\tablenotemark{1}} & \colhead{H\tablenotemark{2}} & \colhead{a\tablenotemark{3}} & \colhead{e\tablenotemark{4}} & \colhead{sin(i)\tablenotemark{5}} &
\colhead{n\tablenotemark{6}} & \colhead{g\tablenotemark{7}} & \colhead{s\tablenotemark{8}}
}
\startdata
326   &    9.29  & 2.3175924 & 0.2033926  & 0.3943312 & 102.025399  &  16.665155  &  -34.712405 \\
1342  &   11.08  & 2.2890380 & 0.2100640  & 0.3636707 & 103.940145  &  19.330543  &  -35.240851 \\
1942  &   13.01  & 2.3183473 & 0.2066104  & 0.3947821 & 101.974747  &  16.658029  &  -34.847479 \\
7703  &   14.59  & 2.3059156 & 0.2007934  & 0.3651971 & 102.800888  &  19.380284  &  -35.289497 \\
16635 &   14.01  & 2.2982526 & 0.2211551  & 0.3972546 & 103.316078  &  16.412431  &  -34.883484 \\
27851 &   13.62  & 2.3102585 & 0.1877882  & 0.3897257 & 102.511556  &  16.847250  &  -33.895959 \\
29475 &   13.60  & 2.3566943 & 0.2122856  & 0.3890748 &  99.496381  &  17.843176  &  -36.794154 \\
31359 &   14.63  & 2.2724503 & 0.2004740  & 0.4035276 & 105.081185  &  15.310253  &  -32.856066 \\
\enddata
\tablecomments{Table~\ref{t:list} is published in its entirety in the machine-readable format.
      A portion is shown here for guidance regarding its form and content.}
\tablenotetext{1}{Asteroid name or designation}
\tablenotetext{2}{Absolute magnitude (mag)}
\tablenotetext{3}{Proper semimajor axis (au)}
\tablenotetext{4}{Proper eccentricity}
\tablenotetext{5}{Sine of proper inclination}
\tablenotetext{6}{Mean motion (deg/yr)}
\tablenotetext{7}{Proper frequency of node (arcsec/yr)}
\tablenotetext{8}{Proper frequency of perihelion (arcsec/yr)}

      \label{t:list}
\end{deluxetable}

\section{The Tamara asteroid family}

\subsection{Size and escape velocity of the parent body}

To gain more insight into the family and its evolution, it is
necessary to estimate the escape velocity from the parent body as well
as the size of its parent body as these two are known to be related
\citep{Sachse2015}.

A simple way to estimate the size of the parent body is to sum up the
diameters of the largest and the third largest family member
\citep{tanga1999}. In this case these are the asteroids (326)~Tamara
and (1942)~Jablunka, with diameters of 89.4 and 16.7~km,
respectively. This gives a diameter for the parent body of
$106.1$~km. Consequently, the largest remnant contains about 60\% of
the total mass of the family, indicating that the Tamara family was
formed in a typical catastrophic collision.

Assuming a density of 1,300~$kg~m^{-3}$, typical for $C$-type
asteroids \citep{carry2012}, we estimate the escape velocity from the
parent body to be 45~m/s.

\subsection{The cumulative distribution of the absolute magnitudes}
\label{ss:mfd}

Another important characteristic of the family is its cumulative
magnitude frequency distribution (CMFD), which should follow a power
law of the form $N(<H) \approx 10^{\alpha H}$. The slope parameter
$\alpha$ can be estimated by numerically fitting the CMFD of the
family members in a specific range of absolute magnitude. Here we
performed this fitting in the $14.5-16.5$ range, and found that
$\alpha = 0.42 \pm 0.02$. This is illustrated in the bottom-right
panel of Figure~\ref{Fig:4panels}. The relatively shallow slope
suggests that the Tamara family is probably not young, as the young
families are typically characterized by somewhat steeper slopes
\citep{vok2006}.

It is interesting to compare the slope of the family and to that of
Phocaeas in general, to estimate how large is the fraction of the
family members among small Phocaeas.  To this end, we derived the
slope of the CMFD of all Phocaeas in the same magnitude range as for
the Tamara family, i.e., $14.5-16.5$~mag, and obtained $\alpha =
0.28\pm0.01$. The slope is significantly shallower than the one
derived for the Tamara family. This is somewhat expected, as the
population of Phocaeas is dominated by the very old Phocaea
collisional family.

Finally, assuming that the derived slopes of the CMFDs are valid for
magnitudes larger than those used to compute them, we predict the
number of all Phocaeas and of Tamara family members to be about 24,500
and 4,200, respectively, for $17<H<20$.  That is, about 17\% of all
Phocaeas should be members of the Tamara family. The fraction of family
members is even larger for magnitudes $H>20$, meaning that among the
small Phocaeas there may be as many dark asteroids as bright ones. See
Section~\ref{ss:neo_flux} for additional discussion on this.

\subsection{Age of the family}
\label{ss:age}

We used the ``V-shape'' method, which is based on the size-dependent
secular drift in semimajor axis induced by the Yarkovsky effect, to
estimate the age of the new family \citep{vok2006,spoto2015}. The existence of such a
structure could also be used to verify the collisional origin of the
group \citep{walsh2013,bolin2017}.

\begin{figure}
\begin{center}
 \includegraphics[width=0.65\textwidth,angle=-90]{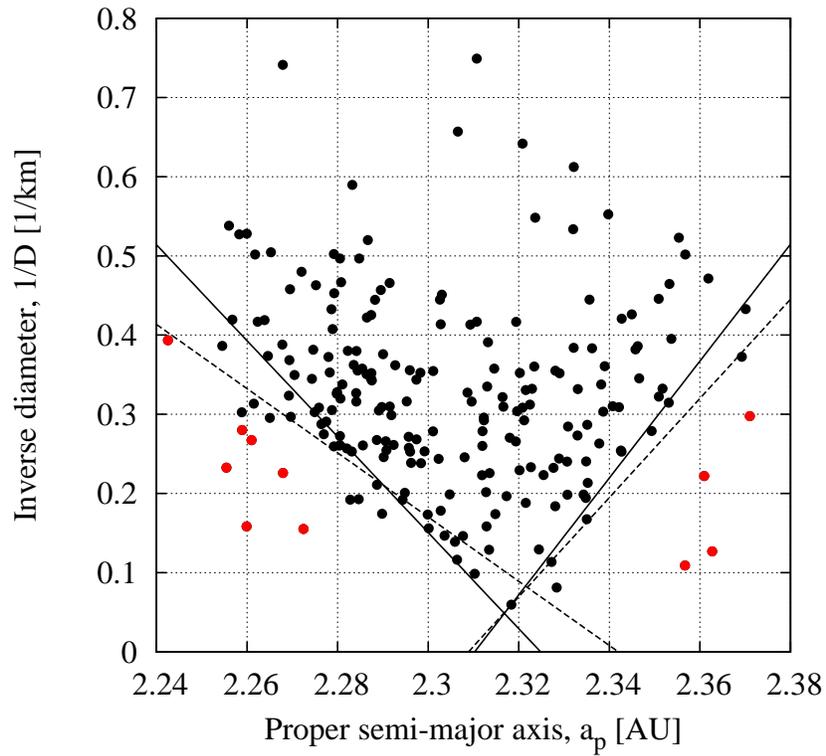}
 \caption{Proper semimajor axis versus the inverse diameter for the
   members of the family.  The dashed and solid lines correspond to
   the initial and final V-shape fit, respectively.  Red dots mark the
   outliers removed from the fit after the first iteration.}
 \label{Fig:v}
 \end{center}
\end{figure}

From the available physical data we computed the mean albedo for
family members to be $\bar{p_v}=0.059\pm0.016$. Using this value we
can convert absolute magnitudes to diameters\footnote{We estimated
  diameters in this way only for the members lacking a direct
  estimate. This is because the infra-red surveys measure emitted
  flux, that is then used to derive the diameters. As a result, the
  diameters obtained are more reliable than the albedos
  \citep{wise0}.} and plot the semimajor axis versus the inverse of
the diameter $(a_p,1/D)$, as shown in Figure~\ref{Fig:v}. The V-shape
structure is clearly visible, providing evidence that the Tamara
family is a real collisional family.

In order to estimate the age of the Tamara family we employed a method
very similar to the one proposed by \citet{spoto2015}. First we
divided the family into left (inner) and right (outer) side with
respect to the barycenter. Then we divided the $1/D$-axis in intervals
containing equal numbers of asteroids, and identified the objects with
the minimum/maximum value of $a_p$ for each interval on the left/right
side.

This data was used to perform a two-step fitting procedure to
determine the slopes of the distribution of the family members in the
$(a_p,1/D)$ plane. We fit the lines through these furthest objects on
both sides in the $(a_p,1/D)$ plane. Then the objects located more
than 0.045 below the lines (in $1/D$) were removed from the
calculation as outliers. Additionally, on the left (inner) side we
also removed a single object with $a_p < 2.25$~au, because its
semimajor axis may be affected by the 7/2 resonance with Jupiter
\citep{ivana2015}. After removing these objects, we again fitted the
members with the minimum/maximum value of $a_p$ for each interval and
obtained the slopes of the V-shape (Figure~\ref{Fig:v}).

The method of family age estimation based on V-shapes requires a
Yarkovsky calibration, that is, an estimate for the maximum value of
the Yarkovsky driven secular drift $(da/dt)_{max}$ for a hypothetical
family member of diameter $D=1$~km \citep{milani2014}.  The value of
$(da/dt)_{max}$ is determined using a model of the Yarkovsky effect
and assuming thermal parameters appropriate for regolith-covered
$C$-type objects \citep{vok2015}.  We adopted $\rho_{s}$ = $\rho_{b}$
= 1300~kg~m$^{-3}$ for the surface and bulk densities
\citep{carry2012}, $\Gamma$ = 250~J~m$^{-2}$~s$^{-1/2}$~K$^{-1}$ for
the surface thermal inertia \citep{delbo2009}, and $\epsilon$ = 0.95
for the thermal emissivity parameter.  With these parameters we
estimated that the maximum drift speed $(da/dt)_{max}$ is about $5.3
\times 10^{-4}$~au/Myr for a body with $D=1$~km.

Finally, using the inverse slopes and the adopted Yarkovsky
calibration we estimated the age of the family to be $264\pm43$~Myr.

\section{Dynamical evolution of the Tamara family}

Numerous gravitational and non-gravitational perturbations constantly
modify asteroid orbits \citep[e.g.][]{nes2015}. Also the group of
asteroids forming an asteroid family evolves in time, being gradually
dispersed by mean motion or secular resonances
\citep{nov2010the,nov2015hof,carruba2016hi}, close encounters with
planets or massive asteroids \citep{carruba2003,nov2010lix} and the
Yarkovsky effect \citep{bottke2001,vok2015}.  Thus, it is of great
importance to reconstruct the dynamical evolution of the family, as
this may help for example to set additional constrains about its age,
or to evaluate the possible leakage from the family towards the
near-Earth region.

\subsection{Dynamical model and initial conditions}

To simulate the dynamical evolution of the Tamara family we performed
a set of numerical integrations.  For this purpose we employed the
\emph{ORBIT9} integrator embedded in the multipurpose \emph{OrbFit}
package\footnote{Available from http://adams.dm.unipi.it/orbfit/}.
The dynamical model includes the gravitational effects of the Sun and
seven major planets, from Venus to Neptune, and also accounts for the
Yarkovsky effect.

Our simulations follow the long-term orbital evolution of test
particles initially distributed randomly inside an ellipse determined
by the Gauss equations. This ellipse corresponds to the dispersion of
the Tamara family members immediately after the breakup event,
assuming an isotropic ejection of the fragments from the parent body \citep{morby1995,nes2002}.
The center of the ellipse coincides with a position of asteroid (323)~Tamara
in the space of osculating semimajor axis, eccentricity and inclination.
At the beginning of the integration, we set the three angular elements (nodal 
longitude, argument of perihelion and mean anomaly) for all test particles
equal to the corresponding elements of asteroid Tamara.
This ensures that Gaussian ellipse defined in the space of
osculating elements corresponds to the position of the Tamara family
in the proper elements space.

The ellipse is obtained assuming a relatively conservative velocity change of 20~m/s
\citep[see e.g.][for typical ejection velocities of asteroid families]{carnes2016}.
Our analysis presented is not very sensitive to this choice. In particular,
estimation of the flux towards the NEO region is not significantly affected by 
the initial size of the family. A larger initial velocity field would, however, cause fragments to reach
the NEO region somewhat earlier than in the case studied here.

For simplicity, the Yarkovsky effect is approximated as a pure
along-track acceleration, inducing on average the same semimajor axis
drift speed $da/dt$ as predicted from theory.

Since the Yarkovsky effect scales as $ \propto 1/D$, the particle
sizes are used to calculate the corresponding value of $(da/dt)$ for
each particle, by scaling from the reference value derived for a
$D=1$~km object (see Section~\ref{ss:age}). Assuming an isotropic
distribution of spin axes in space, to each particle we randomly
assign a value from the $\pm (da/dt)$ interval.

To obtain the sizes of the test particles, we first assign them
absolute magnitude values which follow a CMFD with the same slope as
that of the real family (see Section~\ref{ss:mfd}), and then convert
them to diameters.

\subsection{The dynamical evolution: outcome of numerical simulations}

\begin{figure}
\begin{center}
 \includegraphics[width=0.7\textwidth,angle=-90]{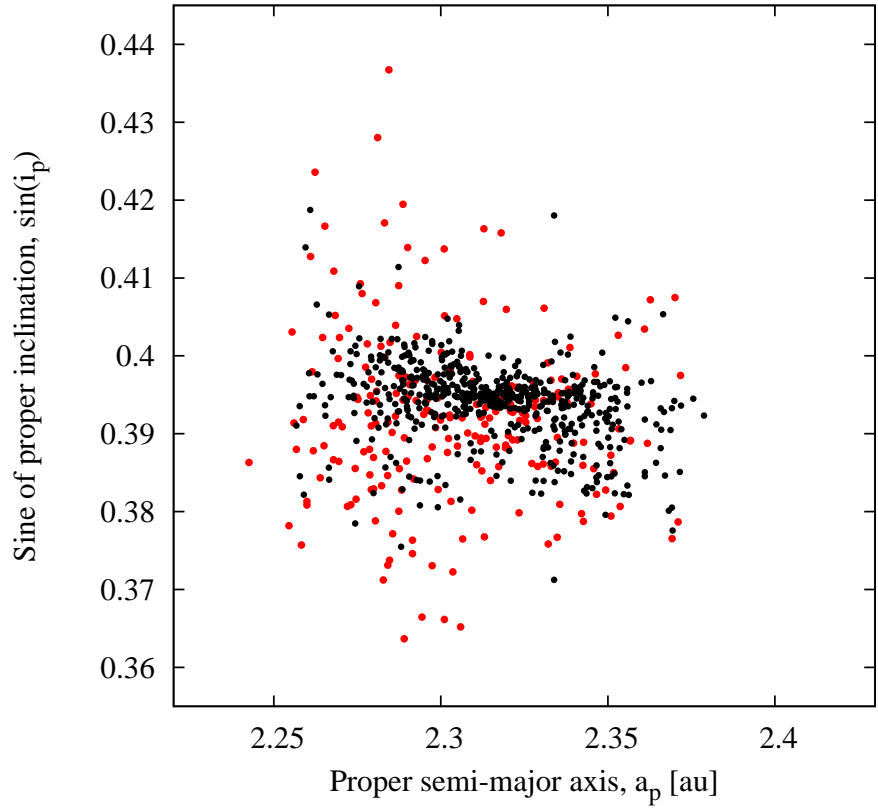}
 \caption{Real and simulated Tamara family in the space of proper orbital semimajor axis vs. sine of proper orbital inclination. The red points represent the real family members, while the black points show the distribution of the test particles after 260 Myr of the evolution. Note that only test particles covering the same size range as the real family members are shown.}
 \label{Fig:famvssim}
 \end{center}
\end{figure}

\begin{figure}
\begin{center}
 \includegraphics[width=1.2\textwidth,angle=-90]{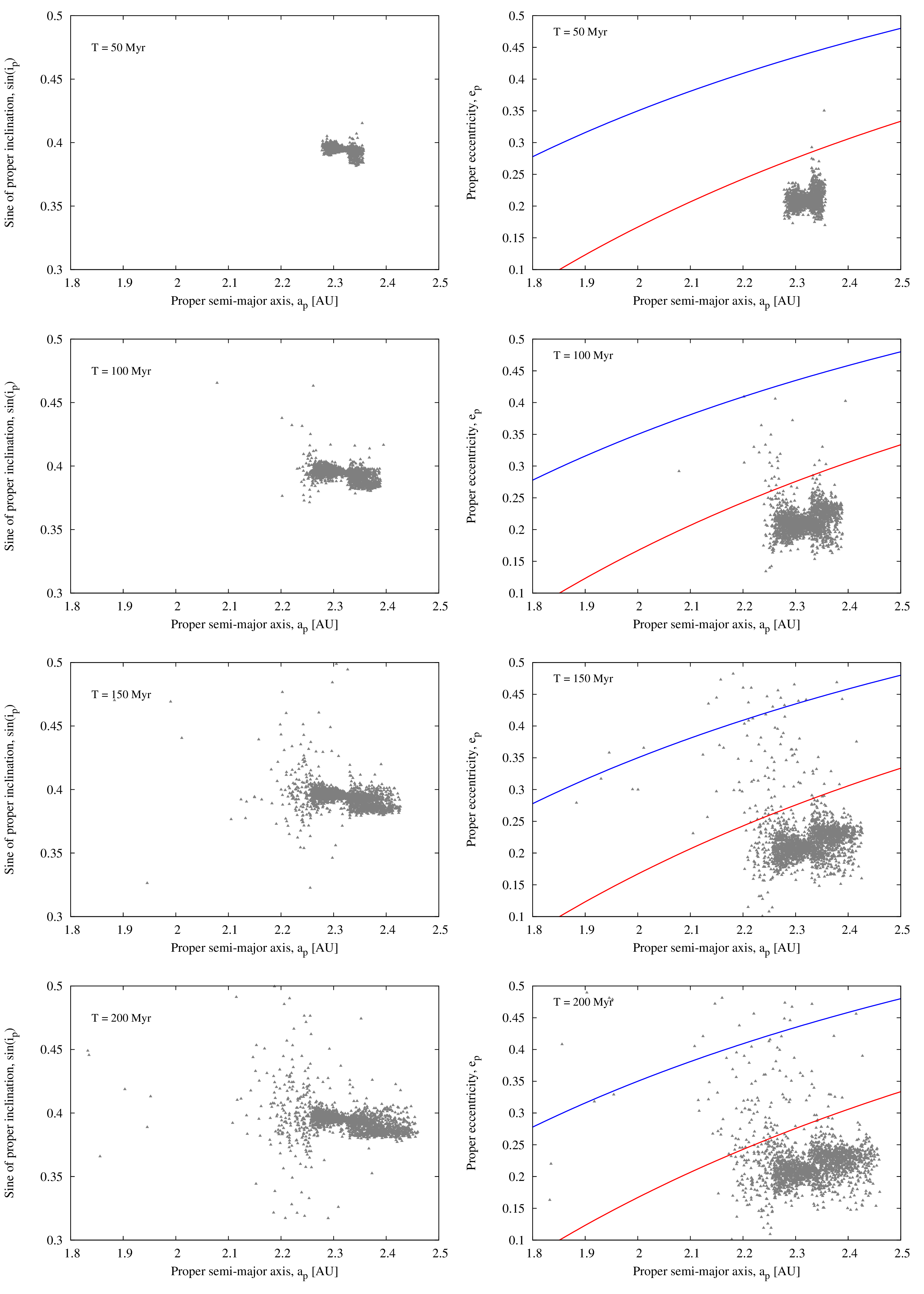}
 \caption{The evolution of the Tamara family in the space of proper
   orbital elements.  The four panels in each column show the
   distribution of the test particles after $50$, $100$, $150$ and
   $200$~Myr (from top to bottom) of evolution. The red and blue lines
   mark perihelion distances $q$ of $1.666$, $1.3$~au, used
   to define populations of Mars-crossers and near-Earth objects, respectively.}
 \label{Fig:evo}
 \end{center}
\end{figure}

The dynamical evolution of the family in the proper elements space was
simulated over $350$~Myr, i.e., about $90$~Myr longer than the
estimated age of the family. The results obtained after $250-300$~Myr
of evolution very nearly matched the current spreading of the family
in the space of proper orbital elements (Figs. \ref{Fig:famvssim} and \ref{Fig:evo}). 
This provides independent
evidence that the Tamara family is indeed the evolutionary outcome of
a fragmentation event in this region.

Yet, a small part of the low-inclination
region of the family was not fully reproduced---specifically around proper semimajor axis of 2.3~au 
(see Fig.~\ref{Fig:famvssim}).
There are several family members located at sine of proper inclination
smaller than 0.37, but this location was not reached by the test particles.
Therefore, we speculate that these objects might have been
injected there due to an anisotropic ejection velocity field, or, they
may not be real family members, but interlopers associated to the
family as a result of the chaining effect, a well known drawback of the HCM 
\citep[see e.g.][]{nov2012}.

Being located at the inner edge of the main asteroid belt, the Tamara
family is potentially an important source of low-albedo near-Earth
objects (NEOs). Since the current members of the family, being large
enough, are still far from the resonances capable of transporting
asteroids close to Earth, only smaller members which drift faster
could have contributed to the NEO flux. Therefore, we focus here on
objects with $17 < H < 19.35$~mag.

\subsubsection{The flux toward the NEO region}
\label{ss:neo_flux}

In order to estimate the number of NEOs originating from the family,
we analyzed the outputs of the integrations, looking for those
particles that at some point over the covered time span, reached
perihelion distances below $1.3$~au. We determined the total number of
objects reaching the near-Earth region as a function of time, as well
as the number of members settled in the NEO space at any specific
point in time.

Based on the CMFD slope $\alpha$ of the Tamara family we estimated that 
there should be 2,280 real members with $17 < H < 19.35$~mag.
We used this number of test particles to study the flux toward the NEO region,
as they should represent the real family in the considered range of magnitudes.

\begin{figure}
\begin{center}
 \includegraphics[width=0.7\textwidth,angle=-90]{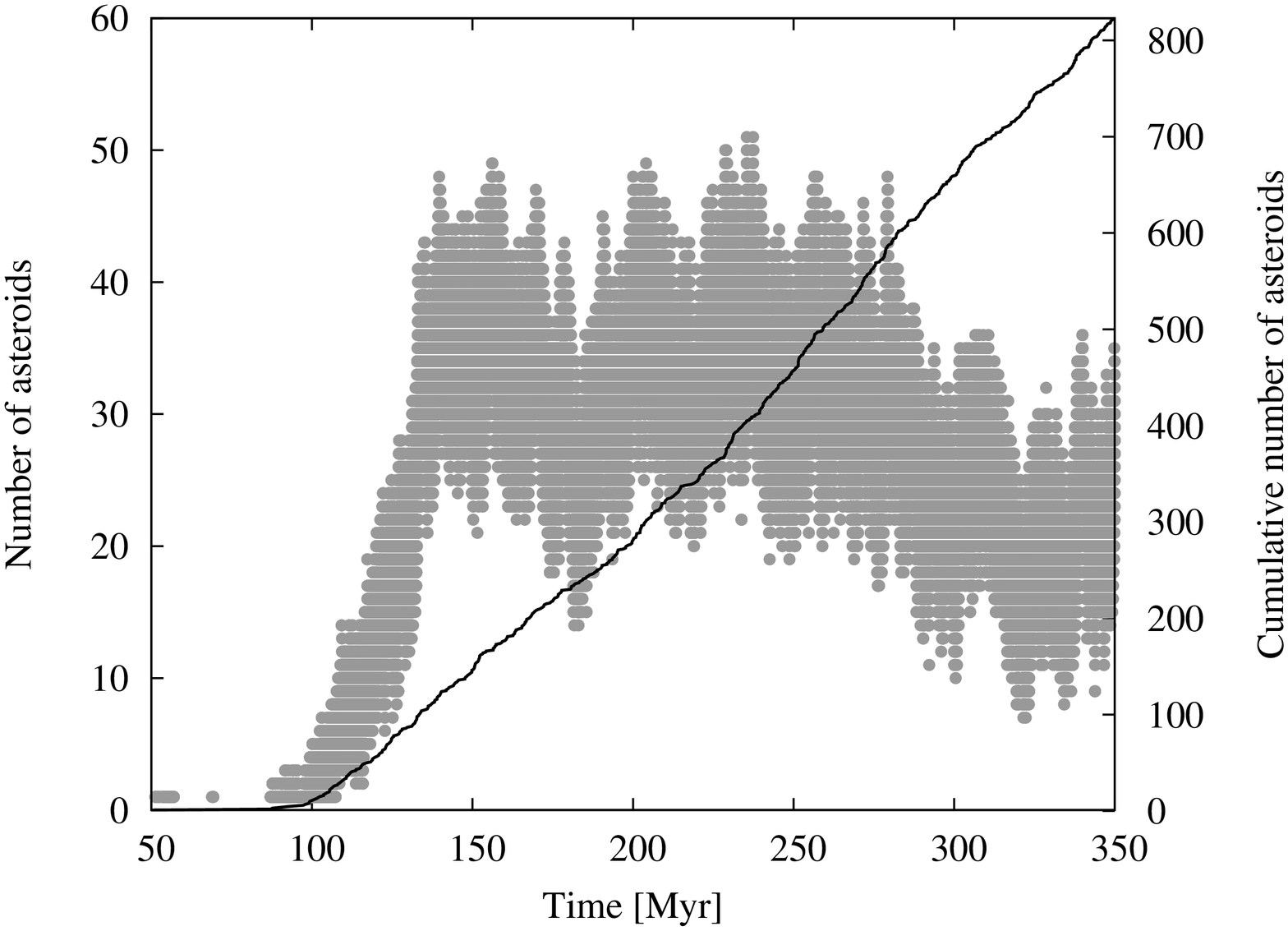}
 \caption{The Tamara asteroid family as a source of NEOs. The
   cumulative number of test particles entering the near-Earth region
   (bold curve), and the total number of members residing in the
   near-Earth region at any specific point in time (gray dots).}
 \label{Fig:neos}
 \end{center}
\end{figure}

The results are shown in Fig.~\ref{Fig:neos}. In this figure the bold
line shows the cumulative number of particles reaching perihelion
distances $q<1.3$~au. It seems that the first family members became
NEOs about 100~Myr after the family formation event, with about 800
test particles in total reaching this area during the subsequent $250$
Myr of the simulation. Therefore, the cumulative number is increasing
almost linearly with time, suggesting that the flux from the family is
about 3 test particles per Myr.  Given the estimated age of the Tamara
family of $264$~Myr, about 500 of its members with $17 < H < 19.35$
should have reached NEO space so far.

In Fig.~\ref{Fig:neos} the gray points represent the number of the
Tamara family members residing in the near-Earth region at any
specific point in time. This result suggests that up to 50 objects
from the family may be found in the region of terrestrial
planets. Focusing on the time interval between $221$ and $307$ Myr of
the evolution (lower and upper limit of the age of the family), we
found that currently there should be $31\pm6$ family members in the
NEO space. Taking into account other possible uncertainties, such as 
the uncertainty of the MFD slope, we found that the maximum 
possible contribution of the family is about $50$ asteroids.

It is interesting to compare this number with the recent model of NEOs
population by \citet{mikael2016a}. These authors found that, in the
magnitude range we considered here, there should be about 250 Phocaeas
residing in the NEOs space.  Hence, about $13\%$ of all NEOs coming
from the Phocaea region should originate in the newly discovered
family, with an upper limit of about $20\%$. 
Moreover, having a somewhat steeper slope of the cumulative magnitude
distribution than the rest of the Phocaeas, the contribution of the
Tamara family is likely to be even larger for smaller objects with
$D<700$~m.

Moreover, the \citet{mikael2016a,mikael2017} NEO model
  predicts that the number of NEOs with $17<H<22$ originating in the
  Phocaea region should be about 670, which represents about $3\%$ of
  the entire NEO population. This is comparable to the fraction of
  objects originating in the outer main belt with $a>3$~au ($3.5\%$)
  and the Jupiter-family comets ($2\%$). The rate of Earth impacts by
  NEOs originating in the Phocaea region is similar to that for NEOs
  from the outer asteroid belt but an order of magnitude greater
  compared to NEOs originating in the Jupiter-family comet population
  (Granvik et al., in preparation). The Tamara family is thus an
  important source for carbonaceous Earth impactors.

\section{Summary and Conclusions}

We present here a detailed study of a population of dark asteroids in
the Phocaea region, and found compelling evidence for the existence of
a new asteroid family within this sub-population of Phocaeas.

We have determined the slope of the cumulative magnitude distribution
of the family, and compared it with the corresponding slope for all
asteroids in the Phocaea region.  This brought us to the conclusion
that for sub-kilometer Phocaeas, the number of dark $C$-type asteroids
is comparable to the number of bright $S$-type objects, questioning
the well established view on this population being almost entirely
composed of rocky asteroids.

Furthermore, based on the standard V-shape method we estimated this
family to be $264\pm43$~Myr old.

Finally, extensive numerical simulations were carried out allowing us
to estimate that $31\pm6$ family members with $H\in [17, 19.35]$,
should exist in the NEO space.

Despite their relatively small number in the near-Earth space,
  the impact rate from small, dark Phocaeas is non-negligible and may
  be an important source for dark meteorites whose parent bodies have
  $17<H<22$. We hypothesize that the peak in the
  distribution of meteor streams \citep{brown2010} at about
  $35-40$~degrees \citep[see Fig. 14S in][]{mikael2016a} may partly be
  produced by parent asteroids coming from the Phocaea region.

\acknowledgments

This work has been supported by the European Union [FP7/2007-2013],
project: STARDUST-The Asteroid and Space Debris Network. BN also
acknowledges support by the Ministry of Education, Science and
Technological Development of the Republic of Serbia, Project
176011. MG was funded by grant \#1299543 from the Academy of Finland.
Numerical simulations were run on the PARADOX-III cluster
hosted by the Scientific Computing Laboratory of the Institute of
Physics Belgrade.

\end{document}